\documentclass[preprint,nofootinbib,onecolumn,11pt]{revtex4}
\usepackage{amsmath}
\usepackage{epsfig}
\usepackage{amssymb}
\usepackage{color}
\usepackage{multirow}
\usepackage{epstopdf}
\usepackage{graphicx}
\usepackage{hyperref}

\def\ffr{$f_{\Xi_b}/f_{\Lambda_b}$}
\newcommand{\fLb}{f_{\Lambda_b}}
\newcommand{\fXib}{f_{\Xi_b}}

\begin{document}

\title{\bf\Large Fragmentation-fraction ratio
$f_{\Xi_b}/f_{\Lambda_b}$ in $b$- and $c$-baryon decays}

\author{Hua-Yu Jiang$^{1}$, Fu-Sheng Yu$^{1,2}$\footnote{Corresponding author}\footnote{ Email: yufsh@lzu.edu.cn} }

\address{
$^1$School of Nuclear Science and Technology,  Lanzhou University,
Lanzhou 730000,  China\\
$^2$Research Center for Hadron and CSR Physics, Lanzhou University
and Institute of Modern Physics of CAS, Lanzhou 730000, China}

\begin{abstract}

We study the ratio of fragmentation fractions, $f_{\Xi_b}/f_{\Lambda_b}$, from the measurement of $\Xi_b^0\to \Xi_c^+\pi^-$ and $\Lambda_b^0\to \Lambda_c^+\pi^-$ with $\Xi_{c}^{+}/\Lambda_{c}^{+}\to p K^-\pi^+$. With the branching fraction $\mathcal{B}(\Xi_c^+\to pK^-\pi^+)=(2.2\pm0.8)\%$ obtained under the U-spin symmetry, the fragmentation ratio is determined as $f_{\Xi_b}/f_{\Lambda_b}$=$0.054\pm0.020$. To reduce the above uncertainties, we suggest to measure the branching fractions of $\Xi_c^+\to p \overline K^{*0}$ and $\Lambda_c^+\to \Sigma^+ K^{*0}$ at  BESIII, Belle(II) and LHCb.

\end{abstract}

\maketitle

\section{Introduction}

Bottom quarks can be produced at the high energy colliders, such
as LHC and Tevatron, and then hadronized into $B$ mesons and
$b$-baryons. The probability of a bottom quark fragments into a
certain weakly decaying $b$-hadron is called the fragmentation
fractions, i.e.
$f_{u,d,s}\equiv \mathcal{B}(b\rightarrow B^-,\overline B^0,
\overline B_s^0)$, $f_{\Lambda_b}\equiv
\mathcal{B}(b\rightarrow\Lambda_b^0)$,
$f_{\Xi_b}\equiv \mathcal{B}(b\rightarrow\Xi_b^0, \Xi_b^-)$ and
$f_{\Omega_b}\equiv \mathcal{B}(b\to \Omega_b^-)$.
As non-perturbative effects, the fragmentation fractions can only
be determined by experiments in some phenomenological approaches.

The $B$-meson fragmentation fractions have been measured by LEP,
Tevatron and LHC with a relatively high precision
\cite{Patrignani:2016xqp,Amhis:2016xyh}. However, the current
understanding of $b$-baryon productions is still a challenge.
The total fragmentation fraction of $b$-baryon is the sum of all
the weakly-decaying $b$-baryons,
\begin{align}\label{eq:bbaryon}
f_{\rm baryon}=f_{\Lambda_b}\left(1+2{f_{\Xi_b}\over
f_{\Lambda_b}}+{f_{\Omega_b}\over
f_{\Lambda_b}}\right)=f_{\Lambda_b}(1+\delta),
\end{align}
where the isospin symmetry is assumed as
$f_{\Xi_b^-}=f_{\Xi_b^0}=f_{\Xi_b}$, and $\delta=2{f_{\Xi_b}\over
f_{\Lambda_b}}+{f_{\Omega_b^-}\over f_{\Lambda_b}}$ is the
correction from $f_{\Lambda_b}$ to $f_{\rm baryon}$.
The averages of the total baryon production fractions are \cite{Amhis:2016xyh}
\begin{align}
f_{\rm baryon}=\left\{
\begin{aligned}
&0.084\pm0.011,~~~~~~ \text{at LEP,}
\\
&0.196\pm0.046,~~~~~~ \text{at Tevatron},
\end{aligned}
\right.
\end{align}
which are inconsistent with each other, and of large
uncertainties.

The total fraction of $b$-baryons has not been determined by LHCb
because of its lack of measurements on $\Xi_{b}^{0,-}$ and
$\Omega_{b}^{-}$.
It has been found that the ratio $f_{\Lambda_{b}}/f_{d}$ depends on
the $p_{\rm T}$ of the final states
\cite{Aaltonen:2008zd,Aaltonen:2008eu,Aaij:2011jp,Aaij:2014jyk}.
At LHCb, the kinematic averaging ratio is \cite{Aaij:2011jp}
%\
\begin{align}
{f_{\Lambda_{b}}\over f_{u}+f_{d}}\Big|_{\rm LHCb}=0.240\pm0.022.
\end{align}
It is required for the information of $f_{\Xi_{b}}$ and
$f_{\Omega_{b}}$to determine the other fragmentation fractions at
LHCb due to the constraint of
\begin{align}
f_{u}+f_{d}+f_{s}+f_{\rm baryon}=1.
\end{align}
Since the production of $\Omega_{b}^{-}$ is suppressed compared to
those of $\Xi_{b}^{0,-}$ by the production of an additional
strange quark, the determination of \ffr~is essential to
understand the productions of $b$-baryons and $B$ mesons.

So far, only Refs. \cite{Voloshin:2015xxa,Hsiao:2015txa} have
predicted the ratio \ffr, both based on the processes of
$\Xi_b^-\to J/\psi\Xi^-$ and $\Lambda_b^0\to J/\psi\Lambda$ with
the data given by CDF and D0. At LHCb, the productions of
$\Lambda_{b}$ and $\Xi_{b}$ have been measured by the
heavy-flavor-conserving process of $\Xi_{b}^{-}\to
\Lambda_{b}^{0}\pi^{-}$ \cite{Aaij:2015yoy}, and the charm-baryon
involving decays of $\Xi_b^0\to\Xi_c^+\pi^-$ and
$\Lambda_b^0\to\Lambda_c^+\pi^-$ via $\Xi_c^+/\Lambda_c^+\to
pK^-\pi^+$ \cite{Aaij:2014esa}.
All the results are listed in Table. \ref{tab:list}.
The production with the charm-baryon involving method is of the
most high precision. The ratio \ffr~can be obtained as long as the
related branching fractions are determined. Among them, the absolute branching fraction of
$\Xi_c^+\to pK^-\pi^+$ has never been measured \cite{Patrignani:2016xqp}, thus is of the largest ambiguity.
In this work, we determine \ffr~with 
$\mathcal{B}(\Xi_c^+\to pK^-\pi^+)$ obtained under the $U$-spin symmetry.

This article is organized as follows. In Sec. II, we introduce the
status of \ffr. The branching fraction of $\Xi_c^+\to pK^-\pi^+$
and \ffr~are obtained in Sec. III and IV, respectively. Sec. V is
the summary.

\begin{table}[!]
\caption{List of measurements related to the fragmentation
fraction ratio $\fXib/\fLb$.}\label{tab:list}
\begin{center}
\begin{tabular}{|c|c|}
\hline\hline
Measurements  & $\fXib/\fLb$
\\\hline
$f_{\Lambda_b}\cdot\mathcal{B}(\Lambda_b^0\to
J/\psi\Lambda)=(5.8\pm0.8)\times10^{-5}$~\cite{Patrignani:2016xqp}
(CDF,D0)  & $0.11\pm0.03$ \cite{Voloshin:2015xxa}
\\
$f_{\Xi_b}\cdot\mathcal{B}(\Xi_b^-\to
J/\psi\Xi^-)=(1.02_{-0.21}^{+0.26})\times10^{-5}$~\cite{Patrignani:2016xqp}
(CDF,D0) & $0.108\pm0.034$~~\cite{Hsiao:2015txa}
\\ \hline
\multirow{2}{*}{$\frac{f_{\Xi_b}}{f_{\Lambda_b}}\cdot\mathcal{B}(\Xi_b^-\to\Lambda_b^0
\pi^-)=(5.7\pm1.8_{-0.9}^{+0.8})\times10^{-4}$~\cite{Aaij:2015yoy}
(LHCb)} &   ~$0.29\pm0.10$ ~(MIT bag
model)~~\cite{Cheng:2015ckx}\\
 &~$0.08\pm0.03$ ~(diquark model)~~\cite{Cheng:2015ckx}
\\ \hline
%\multirow{2}{*}{$\frac{f_{\Xi_b}}{f_{\Lambda_b}}\frac{\mathcal{B}(\Xi_b^0\rightarrow\Lambda
%\pi^+\pi^-)}{\mathcal{B}(\Lambda_b^0\rightarrow(\Lambda\pi^+)_{\Lambda_c^+}\pi^-)}=(2.0\pm1.0\pm0.8)\times10^{-2}$~\cite{Aaij:2016nrq}}
%&  \multirow{2}{*}{-- --}
%\\
%&
%\\ \hline
\multirow{2}{*}{$\frac{f_{\Xi_b}}{f_{\Lambda_b}}\cdot\frac{\mathcal{B}(\Xi_b^0\to\Xi_c^+
\pi^-)}{\mathcal{B}(\Lambda_b^0\to\Lambda_c^+\pi^-)}\cdot\frac{\mathcal{B}
(\Xi_c^+\to pK^-\pi^+)}{\mathcal{B}(\Lambda_c^+\to pK^-\pi^+)}=
(1.88\pm0.04\pm0.03)\times10^{-2}$~\cite{Aaij:2014esa} (LHCb)} &
\multirow{2}{*}{$0.054\pm0.020$ ~(this work)}
\\
&
\\ \hline\hline
\end{tabular}
\end{center}
\end{table}

%%%%%%%%%%%%%%%%%%%%%%%%%%%%%
\section{Status of $f_{\Xi_b}/f_{\Lambda_b}$}

In some literatures, it is usually assumed that the difference between the
productions of $\Xi_b$ and $\Lambda_b$ is from the strange
quark and up or down quarks
\cite{HerediaDeLaCruz:2011yi,Aaij:2014esa},
\begin{align}\label{eq:fsfu}
{f_{\Xi_b}\over f_{\Lambda_b}}\simeq {f_s\over
f_u},~~~~~~\text{or}~~~~~~{f_{\Xi_b}\over f_{\Lambda_b}}\simeq
0.2.
\end{align}
However, since the fragmentation fractions are non-perturbative
effects, they can only be extracted from experimental data. In
this section, we introduce the status of $f_{\Xi_b}/f_{\Lambda_b}$
by means of the relevant measurements.

\subsection{$\Xi_b^-\to J/\psi\Xi^-$ v.s. $\Lambda_b^0\to
J/\psi\Lambda$}

So far, the only theoretical analysis on \ffr~are performed in Refs.
\cite{Voloshin:2015xxa,Hsiao:2015txa} based on the experimental
data of $\Xi_b^-\to J/\psi\Xi^-$ and $\Lambda_b^0\to
J/\psi\Lambda$. In Ref. \cite{Patrignani:2016xqp}, the relevant results averaging the measurements by CDF and D0 \cite{Aaltonen:2009ny,Abazov:2007am,Abazov:2011wt,Abe:1996tr}, are
given as
\begin{align}\label{eq:Jpsi}
\begin{split}
&f_{\Lambda_b}\cdot\mathcal{B}(\Lambda_b^0\to
J/\psi\Lambda)=(5.8\pm0.8)\times10^{-5},
\\
&f_{\Xi_b}\cdot\mathcal{B}(\Xi_b^-\to
J/\psi\Xi^-)=(1.02_{-0.21}^{+0.26})\times10^{-5}.
\end{split}
\end{align}
The fragmentation fraction ratio of \ffr~can be obtained unless
the ratio of branching fractions of $\Xi_b^-\to J/\psi\Xi^-$ and
$\Lambda_b^0\to J/\psi\Lambda$ is known.

Both $\Xi_b^-\to J/\psi\Xi^-$ and $\Lambda_b^0\to J/\psi\Lambda$
are the $b\to c \bar c s$ transitions with the spectators of
$(ds-sd)/\sqrt2$ and $(ud-du)/\sqrt2$, respectively. Therefore,
the two processes are related to each other under the flavor
$SU(3)$ symmetry. The width relation of
\begin{align}
\Gamma(\Lambda_b^{0}\to
J/\Psi\Lambda)=\frac{2}{3}\Gamma(\Xi_b^-\to J/\Psi \Xi^-),
\end{align}
is given by Voloshin \cite{Voloshin:2015xxa}. Using the
experimental data in (\ref{eq:Jpsi}), the ratio of the
fragmentation fractions can then be obtained as
\cite{Voloshin:2015xxa}
\begin{align}\label{eq:voloshin}
\frac{f_{\Xi_b}}{f_{\Lambda_b}}=0.11\pm0.03.
\end{align}

Hsiao {\it et ac} express the decay amplitudes of  $\Xi_b^-\to
J/\psi\Xi^-$ and $\Lambda_b^0\to J/\psi\Lambda$ in the
factorization approach \cite{Hsiao:2015txa}. They relate the form
factors of $b$-baryon to light baryon octet transitions based on
the $SU(3)$ symmetry, and obtain the ratio of branching fractions
$\mathcal{B}(\Xi_b^-\to J/\psi\Xi^-)/\mathcal{B}(\Lambda_b^0\to
J/\psi\Lambda)=1.63\pm0.04$. Utilizing the data in
Eq. \eqref{eq:Jpsi}, the authors give a result similar to
Eq. (\ref{eq:voloshin}),
\begin{align}\label{eq:hsiao}
\frac{f_{\Xi_b}}{f_{\Lambda_b}}=0.108\pm0.034.
\end{align}

\subsection{Heavy-flavor-conserving decay
$\Xi_b^-\to\Lambda_b^0\pi^-$}

The LHCb collaboration has observed the first
heavy-flavor-conserving $\Delta S=1$ hadronic weak decay
$\Xi_b^-\to\Lambda_b^0\pi^-$ \cite{Aaij:2015yoy}, with
\begin{align}\label{eq:StrangeWD}
\frac{f_{\Xi_b}}{f_{\Lambda_b}}\cdot\mathcal{B}(\Xi_b^-\to\Lambda_b^0
\pi^-)=(5.7\pm1.8_{-0.9}^{+0.8})\times10^{-4}.
\end{align}
 In Ref. \cite{Aaij:2015yoy}, $f_{\Xi_b}/f_{\Lambda_b}$ is assumed to
 be bounded between 0.1 and 0.3, and then obtain the branching
 fraction of $\Xi_b^-\to\Lambda_b^0 \pi^-$ lie in the range from
 $(0.57\pm0.21)\%$ to $(0.19\pm0.07)\%$. On the contrary, the
 fragmentation ratio  \ffr~can be obtained if
 $\mathcal{B}(\Xi_b^-\to\Lambda_b^0\pi^-)$ is determined.

In Ref. \cite{Cheng:2015ckx}, the branching fraction of
$\Xi_b^-\to\Lambda_b^0 \pi^-$ is calculated in the MIT bag model
and the diquark model,
\begin{equation}\label{BRstrangeWD}
\mathcal{B}(\Xi_b^-\to\Lambda_b^0\pi^-)=\left\{
\begin{aligned}
&2.0\times10^{-3},\qquad \text{MIT bag model,}\\
&6.9\times10^{-3},\qquad \text{diquark model.}
\end{aligned}
\right.
\end{equation}
Subsequently, we can obtain the ratio of fragmentation fractions
according to Eq.\eqref{eq:StrangeWD} as,
\begin{align}
 \frac{f_{\Xi_b^-}}{f_{\Lambda_b^0}}=
&0.29\pm0.10,\qquad\qquad \text{MIT bag model,}\label{eq:MIT}\\
 \frac{f_{\Xi_b^-}}{f_{\Lambda_b^0}}= &0.08\pm0.03,\qquad\qquad
 \text{diquark model.}\label{eq:diquark}
\end{align}

\subsection{$\Xi_b^0\to\Xi_c^+\pi^-$ v.s.
$\Lambda_b^0\to\Lambda_c^+\pi^-$ via $\Xi_c^+/\Lambda_c^+\to
pK^-\pi^+$}

In the above two methods, the experimental measurements are of
large uncertainties, as seen in Eqs. \eqref{eq:Jpsi} and
\eqref{eq:StrangeWD}.
In the decay of $\Xi_{b}^{-}\to J/\Psi \Xi^{-}$, the efficiency of
reconstruction of $\Xi^{-}$ with $\Xi^{-}\to \Lambda\pi^{-}$ and
$\Lambda\to p \pi^{-}$, is very small in the hadron colliders
\cite{Abazov:2007am,Aaltonen:2009ny}.
On the other hand, the branching fraction of $\Xi_{b}^{-}\to
\Lambda_{b}^{0}\pi^{-}$ is expected to be very small.

Compared to the above processes, the relative production ratio
between $\Xi_b^0\to\Xi_c^+\pi^-$ and
$\Lambda_b^0\to\Lambda_c^+\pi^-$ has been measured by LHCb with
much higher precision \cite{Aaij:2014esa},
\begin{align}\label{eq:LHCbff}
\frac{f_{\Xi_b^0}}{f_{\Lambda_b^0}}\cdot\frac{\mathcal{B}(\Xi_b^0\to\Xi_c^+
\pi^-)}{\mathcal{B}(\Lambda_b^0\to\Lambda_c^+\pi^-)}\cdot\frac{\mathcal{B}
(\Xi_c^+\to pK^-\pi^+)}{\mathcal{B}(\Lambda_c^+\to pK^-\pi^+)}=
(1.88\pm0.04\pm0.03)\times10^{-2}.
\end{align}
As long as the branching fractions of the relevant $b$- and
$c$-baryon decays are known, Eq. \eqref{eq:LHCbff} could provide a
good determination of \ffr.
In Ref. \cite{Aaij:2014esa}, with naively expected values of
$\mathcal{B}(\Xi_b^0\to\Xi_c^+\pi^-)/\mathcal{B}(\Lambda_b^0\to\Lambda_c^+\pi^-)\approx1$
and $\mathcal{B}(\Xi_c^+\to pK^-\pi^+)/\mathcal{B}(\Lambda_c^+\to
pK^-\pi^+)\approx0.1$, it can be obtained that
$f_{\Xi_b}/f_{\Lambda_b}\approx0.2$.

The branching fraction of $\Xi_b^0\to\Xi_c^+\pi^-$ has not been
directly measured in experiment.
$\mathcal{B}(\Xi_b^0\to\Xi_c^+\pi^-)$ and
$\mathcal{B}(\Lambda_b^0\to\Lambda_c^+\pi^-)$ are equal to each
other in the heavy quark limit and the flavor $SU(3)$ symmetry.
In literatures, only Refs. \cite{Cheng:1996cs} and
\cite{Ivanov:1997hi} calculate both the branching fractions of
$\Xi_b^0\to\Xi_c^+\pi^-$ and $\Lambda_b^0\to\Lambda_c^+\pi^-$.
With the transition form factors in the non-relativistic quark
model, the ratio of branching fractions involving the factorizable
contribution can be obtained in \cite{Cheng:1996cs}:
\begin{align}
\frac{\mathcal{B}(\Xi_b^0\to\Xi_c^+\pi^-)}{\mathcal{B}(\Lambda_b^0\to\Lambda_c^+\pi^-)}
=\frac{\Gamma(\Xi_b^0\to\Xi_c^+\pi^-)}{\Gamma(\Lambda_b^0\to\Lambda_c^+\pi^-)}
=\frac{0.33a_1^2\times 10^{10}s^{-1}}{0.31a_1^2\times
10^{10}s^{-1}}=1.07,
\end{align}
where the difference in the lifetimes is neglected since
$\tau(\Xi_b^0)/\tau(\Lambda_b^0)=1.006\pm0.021$, and
$a_{1}=C_{1}+C_{2}/3$ is the effective Wilson coefficient. The
deviation from unity results from the mass difference between
$m_{\Xi_{b}}+m_{\Xi_{c}}$ and $m_{\Lambda_{b}}+m_{\Lambda_{c}}$,
i.e. the $SU(3)$ breaking effect. In the soft-collinear effective
theory, the non-factorizable contributions from the
color-commensurate and the $W$-exchange diagrams are suppressed by
$\mathcal{O}(\Lambda_{\rm QCD}/m_{b})$ \cite{Leibovich:2003tw}.
In Ref. \cite{Ivanov:1997hi} in a relativistic three-quark model,
it is found that the non-factorizable contributions amount up to
30\% of the factorizable ones, with the ratio of
$
\mathcal{B}(\Xi_b^0\to\Xi_c^+\pi^-)/\mathcal{B}(\Lambda_b^0\to\Lambda_c^+\pi^-)=1.25.
$
Therefore, even without a reliable study in a QCD-based approach,
it can still be expected that the deviation of the ratio from
unity is under control.

The absolute branching fraction of $\Lambda_c^+\to pK^-\pi^+$ has
been well measured by Belle and BESIII
\cite{Zupanc:2013iki,Ablikim:2015flg}, with $\mathcal{B}(\Lambda_c^+\to
pK^-\pi^+)=(6.35\pm0.33)\%$ \cite{Patrignani:2016xqp}. However, branching fraction
of $\Xi_c^+\to pK^-\pi^+$ is of large ambiguity. The ratio of
$\mathcal{B}(\Xi_c^+\to pK^-\pi^+)/\mathcal{B}(\Lambda_c^+\to
pK^-\pi^+)\approx0.1$ used in \cite{Aaij:2014esa}, is only naively
assumed by the Cabibbo factor. In the next section, we will obtain
the branching fraction of $\Xi_c^+\to pK^-\pi^+$ via $U$-spin
analysis, and then determine \ffr.

\section{Branching fraction of $\Xi_c^+\to pK^-\pi^+$}

The understanding of charmed baryon decays are still of high
deficiency both in theory and in experiment.
So far, there has not been any measurement on the absolute
branching fraction of $\Xi_{c}^{0,+}$ decays
\cite{Patrignani:2016xqp}. The ratio of $\mathcal{B}(\Xi_c^+\to
pK^-\pi^+)/\mathcal{B}(\Xi_c^+\to \Xi^-\pi^+\pi^+)$ has been
measured to be $0.21\pm0.04$
\cite{Link:2001rn,VazquezJauregui:2008eg,Patrignani:2016xqp}. But
it is still unknown for the absolute branching fraction of $\Xi_c^+\to
pK^-\pi^+$.

It is first found in Ref. \cite{Yu:2017zst} that
$\mathcal{B}(\Xi_c^+\to pK^-\pi^+)=(2.2\pm0.8)\%$ can be obtained
from the measured ratio of $\mathcal{B}(\Xi_c^+\to p\overline
K^{\ast0})/\mathcal{B}(\Xi_c^+\to pK^-\pi^+)=0.54\pm0.10$
\cite{Link:2001rn}, and the $U$-spin relation between $\Xi_c^+\to
p\overline K^{\ast0}$ and $\Lambda_c^+\to \Sigma^+K^{\ast0}$. We
show the $U$-spin analysis in detail in the present work.

%There is no absolute branching fraction of $\Xi_c^+\to pK^-\pi^+$
%was measured by experiment, but the ratios of branching fractions
%and $\mathcal{B}(\Xi_c^+\to p\overline
%K^{\ast0})/\mathcal{B}(\Xi_c^+\to pK^-\pi^+)$ were measured with
%the value $(0.21\pm0.04)$ and $(0.54\pm0.10)$, respectively. And
%the two amplitude $\mathcal{A}(\Xi_c^+\to p\bar K^{\ast0})$ and
%$\mathcal{A}(\Lambda_c^+\to \Sigma^+K^{\ast0})$ can be related by
%$U$-spin symmetry, where the branching fraction of the second one
%has been measured as $(0.36\pm0.10)\%$. Firstly, let's derive the
%$U$-spin relation of the two channel. The effective Hamiltonian
%for the weak decay of $c\to d\bar du/s\bar su$ can be expressed as

\begin{figure}[!]
  \centering
  \includegraphics[width=0.7\textwidth]{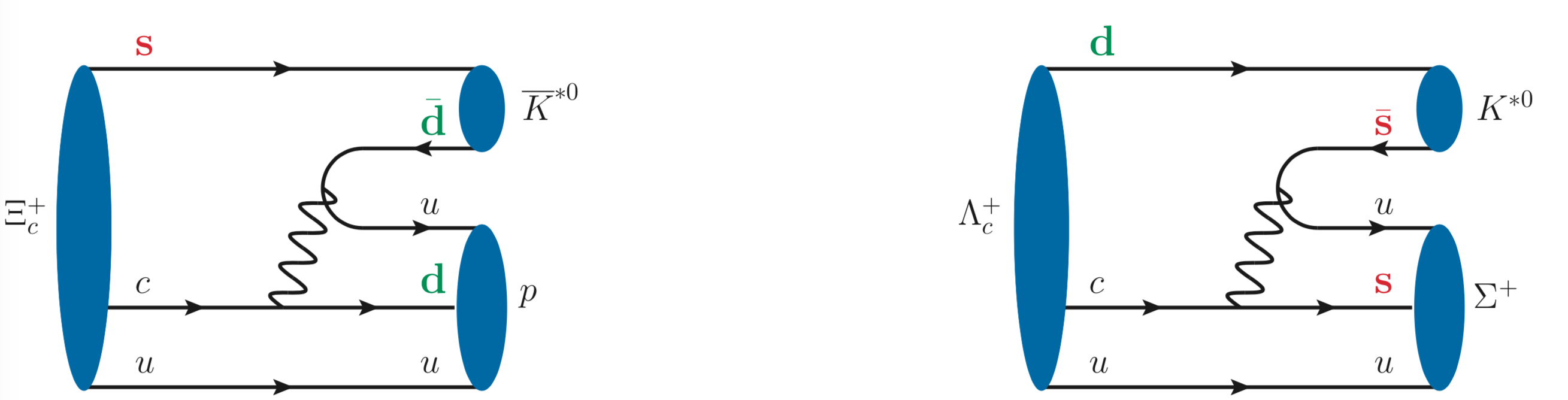}\\
  \includegraphics[width=0.7\textwidth]{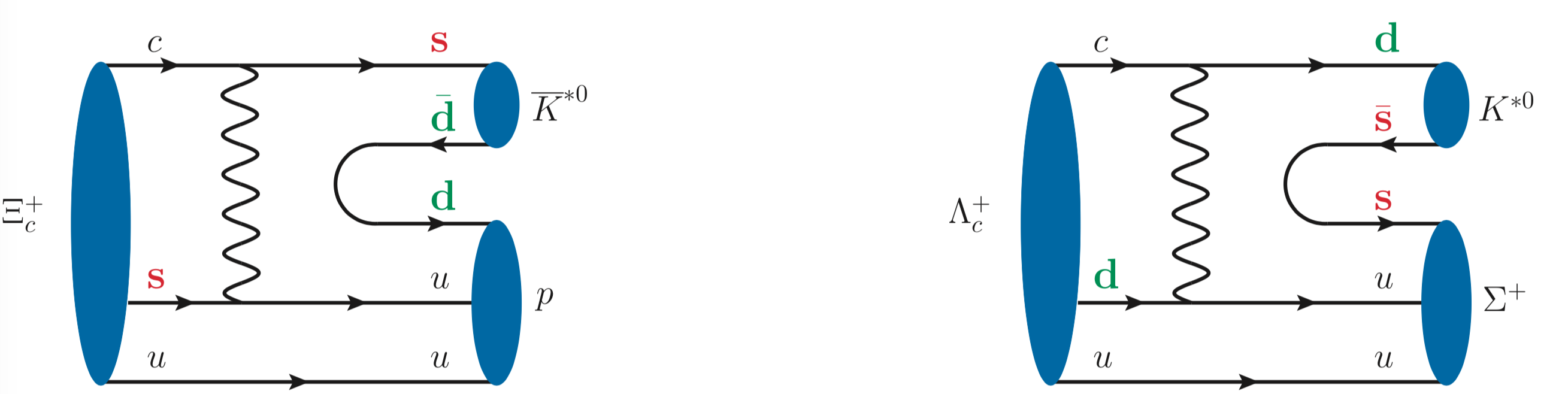}\\
  \caption{The topological diagrams of $\Xi_c^+\to p\overline
  K^{\ast0}$ and $\Lambda_c^+\to \Sigma^+K^{\ast0}$. The top and
  bottom diagrams are color-commensurate and $W$-exchange ones,
  respectively.  }\label{Fig:UspC}
\end{figure}

The decays of $\Xi_c^+\to p\overline K^{\ast0}$ and
$\Lambda_c^+\to \Sigma^+K^{\ast0}$ are both singly
Cabibbo-suppressed modes, with the transition of $c\to (s\bar s-
d\bar d)u$ where the minus sign between $s\bar s$ and $d\bar d$
comes from the Cabibbo-Kobayashi-Maskawa matrix elements,
$V_{cd}^{*}V_{ud}=-V_{cs}^{*}V_{us}$. Note that the $U$-spin
doublets are $(|d\rangle, |s\rangle)$ and $(|\bar s\rangle, -
|\bar d\rangle)$. The effective Hamiltonian of $c\to (s\bar s-
d\bar d)u$ changes the $U$-spin by $\Delta U=1$, $\Delta U_{3}=0$,
i.e.
$|\mathcal{H}_{\rm eff}\rangle=\sqrt{2}|1,0\rangle$. $\Xi_c^+$ and
$\Lambda_c^+$ form a $U$-spin doublet of $(\Lambda_{c}^{+},
\Xi_{c}^{+})$. We have
\begin{align}
\mathcal{H}_{\rm eff}|\Xi_{c}^{+}\rangle
=&\sqrt{2}\left|1,0;{1\over2},-{1\over2}\right\rangle={2\over\sqrt3}\left|{3\over2},-{1\over2}\right\rangle+\sqrt{2\over3}\left|{1\over2},-{1\over2}\right\rangle,
\\
\mathcal{H}_{\rm eff}|\Lambda_{c}^{+}\rangle
=&\sqrt{2}\left|1,0;{1\over2},{1\over2}\right\rangle={2\over\sqrt3}\left|{3\over2},{1\over2}\right\rangle
-\sqrt{2\over3}\left|{1\over2},{1\over2}\right\rangle.
\end{align}
The $U$-spin representations of the $|p\overline K^{\ast0}\rangle$
and $|\Sigma^+K^{\ast0}\rangle$ states are
\begin{align}\label{eq:XipKs}
&\big|p\overline K^{\ast0}\big\rangle
=\left|{1\over2},{1\over2};1,-1\right\rangle
={1\over\sqrt3}\left|{3\over2},-{1\over2}\right\rangle-\sqrt{2\over3}\left|{1\over2},
-{1\over2}\right\rangle,
\\
&\left|\Sigma^+K^{\ast0}\right\rangle=\left|{1\over2},-{1\over2};1,1\right\rangle
={1\over\sqrt3}\left|{3\over2},{1\over2}\right\rangle+\sqrt{2\over3}\left|{1\over2},
{1\over2}\right\rangle.
\end{align}
The decay amplitudes are then
\begin{align}\label{eq:LaSigmaKs}
\mathcal{A}(\Xi_c^+\to p\overline K^{\ast0})
=&\big\langle p\overline K^{\ast0}\big|\mathcal{H}_{\rm
eff}\big|\Xi_c^+\big\rangle
={2\over3}A_{3/2}-{2\over3}A_{1/2},
\\
\mathcal{A}(\Lambda_c^+\to\Sigma^+K^{\ast0})
=&\big\langle\Sigma^+K^{\ast0}\big|\mathcal{H}_{\rm
eff}\big|\Lambda_c^+\big\rangle
={2\over3}A_{3/2}-{2\over3}A_{1/2},
\end{align}
where $A_{3/2}$ and $A_{1/2}$ are the amplitudes of  $U$-spin of
$3/2$ and $1/2$, respectively.
It is clear that the amplitudes satisfy
\begin{align}\label{eq:USrelation}
\mathcal{A}(\Xi_c^+\to p\overline
K^{\ast0})=\mathcal{A}(\Lambda_c^+\to \Sigma^+K^{\ast0}).
\end{align}
This relation can also be seen from the topological diagrams in
FIG.\ref{Fig:UspC}.

According to the relation in Eq.\eqref{eq:USrelation}, the
branching ratio of $\Xi_c^+\to p\overline K^{\ast0}$ can be
obtained by
\begin{align}\label{eq:ratio}
\mathcal{B}(\Xi_{c}^{+}\to p\overline
K^{\ast0})=\frac{m_{\Lambda_c^+}^2}{m_{\Xi_c^+}^2}\cdot
\frac{\tau_{\Xi_c^+}}{\tau_{\Lambda_c^+}}\cdot
\frac{|p_c(m_{\Xi_c},m_{p},m_{K^{*}})|}{|p_c(m_{\Lambda_c},m_{\Sigma},m_{K^{*}})|}\cdot
\mathcal{B}(\Lambda_c^+\to\Sigma^+K^{\ast0}),
\end{align}
where
$|p_c(M,m_{1},m_{2})|={\sqrt{[M^2-(m_1+m_2)^2][M^2-(m_1-m_2)^2]}}/{2M}$.
The data of masses and lifetimes are taken from PDG
\cite{Patrignani:2016xqp}: $m_{\Xi_{c}}=2468$MeV,
$m_{\Lambda_{c}}=2286$ MeV, $m_{p}=938$ MeV, $m_{\Sigma}=1189$
MeV, $m_{K^{*}}=892$ MeV,
$\tau_{\Xi_{c}^{+}}=(4.42\pm0.26)\times10^{-13}$ s,
$\tau_{\Lambda_{c}^{+}}=(2.00\pm0.06)\times10^{-13}$ s. Besides, 
$\mathcal{B}(\Lambda_c^+\to\Sigma^+\pi^{+}\pi^{-})=(4.57\pm0.29)\%$
\cite{Patrignani:2016xqp,Ablikim:2015flg}, and the branching ratios are \cite{Link:2002zx,Link:2001rn}
\begin{align}
{\mathcal{B}(\Lambda_c^+\to\Sigma^+K^{\ast0})\over \mathcal{B}(\Lambda_c^+\to\Sigma^+\pi^{+}\pi^{-})}&=0.078\pm0.022, \label{eq:Br1}
\\
{\mathcal{B}(\Xi_c^+\to
p\overline K^{\ast0})\over\mathcal{B}(\Xi_c^+\to
pK^-\pi^+)}&=0.54\pm0.10 .\label{eq:Br2}
\end{align}
 Then we can obtain
\begin{align}\label{eq:Xic}
\mathcal{B}(\Xi_c^+\to pK^-\pi^+)=(2.2\pm0.8)\%.
\end{align}
The uncertainty is dominated by the ratios of branching fractions
of $\Lambda_{c}^{+}$ and $\Xi_{c}^{+}$ decays in Eqs. \eqref{eq:Br1} and \eqref{eq:Br2}. 
%This is the first prediction on the branching fraction of $\Xi_c^+\to pK^-\pi^+$.

The central value of $\mathcal{B}(\Xi_c^+\to pK^-\pi^+)$ at the
order of percent, is larger than the typical order of $10^{-3}$ of
the
ordinary singly Cabibbo-suppressed processes, such as
$\mathcal{B}(\Lambda_c^+\to\Sigma^+K^{\ast0})=(3.6\pm1.0)\times10^{-3}$.
This can be clarified by Eq. (\ref{eq:ratio}). Firstly, the
lifetime of $\Xi_{c}^{+}$ is two times larger than that of
$\Lambda_{c}^{+}$. Secondly, the phase space of $\Xi_{c}^{+}\to
p\overline K^{\ast0}$ is larger than that of
$\Lambda_c^+\to\Sigma^+K^{\ast0}$ by another factor of two, i.e.
$|p_{c}(m_{\Xi_c},m_{p},m_{K^{*}})|=828$ MeV and
$|p_c(m_{\Lambda_c},m_{\Sigma},m_{K^{*}})|=470$ MeV. Due to the
larger lifetime and phase space, the branching fraction of
$\Xi_{c}^{+}\to p\overline K^{\ast0}$ is then at the order of
percent, $(1.2\pm0.4)\%$.

The understanding of the dynamics of charmed baryon decays is
still a challenge at the current stage. Recent theoretical studies
are mostly based on the flavor $SU(3)$ analysis
\cite{Lu:2016ogy,Wang:2017gxe,Geng:2018plk,Geng:2017mxn,Geng:2017esc}
and the current algebra \cite{Cheng:2018hwl}. They have not yet
been applied to the singly Cabibbo-suppressed charmed baryon
decays into a light baryon and a vector meson. Therefore, it is
not available to estimate the $U$-spin breaking effects in the
above analysis of Eq. \eqref{eq:Xic}. In the $D$-meson decays,
the $U$-spin breaking effects, or say the $SU(3)$ breaking
effects, are mainly from the transition form factors and decay
constants in the factorizable amplitudes, the difference between
$u\bar u$, $d\bar d$ and $s\bar s$ produced from vacuum in the
$W$-exchange and $W$-annihilation amplitudes, and the Glauber
strong phase with pion in the non-factorizable
contributions \cite{Li:2012cfa,Li:2013xsa}. In Fig. \ref{Fig:UspC},
both amplitudes in the $\Xi_{c}^{+}\to p\overline K^{*0}$ and
$\Lambda_{c}^{+}\to \Sigma^{+}K^{*0}$ decay are non-factorizable.
The vacuum production of $d\bar d$ and $s\bar s$ in the
$W$-exchange diagrams would be a main source of $U$-spin breaking.
In the modes involving a vector meson and a pseudoscalar meson in
the final states of $D$-meson decays, the difference between
$d\bar d$ and $s\bar s$ production in the $W$-exchange diagrams
can be seen from $\chi_{d}^{E}e^{i \phi_{d}^{E}}=(0.49\pm0.03)e^{i
(92\pm4)^{\circ}}$ and $\chi_{s}^{E}e^{i
\phi_{s}^{E}}=(0.54\pm0.03)e^{i (128\pm5)^{\circ}}$
\cite{Jiang:2017zwr} where $\chi$ and $\phi$ are the magnitude and
strong phase of the non-perturbative parameters in the
$W$-exchange diagrams, and the subscripts $d$ and $s$ denotes the
quark flavor of $q\bar q$ produced from the vacuum. It can be found that the $U$-spin
breaking effects are not large in $D\to VP$ modes. The
$W$-exchange diagrams in charmed baryon decays are similar to the
ones in charmed meson decays, with an additional spectator quark.
It can be expected that the $U$-spin breaking effects between
$\Xi_{c}^{+}\to p\overline K^{*0}$ and $\Lambda_{c}^{+}\to
\Sigma^{+}K^{*0}$ would not be large, and thus the prediction of $\mathcal{B}(\Xi_c^+\to pK^-\pi^+)=(2.2\pm0.8)\%$ would be under control.

The process of $\Xi_c^+\to pK^-\pi^+$ with all the charged final
particles is widely used to study the properties of, or to search
for some heavier baryons. The mass and lifetime of $\Xi_{b}^{0}$
are measured with the most high precision via $\Xi_{b}^{0}\to \Xi_{c}^{+}\pi^{-}$, $\Xi_c^+\to
pK^-\pi^+$ \cite{Aaij:2014esa}. New states of
$\Xi_{b}^{\prime-}({1\over2}^{+})$ and
$\Xi_{b}^{*-}({3\over2}^{+})$ are observed in the
$\Xi_{b}^{0}\pi^{-}$ spectrum with $\Xi_{b}^{0}\to
\Xi_{c}^{+}\pi^{-}$, $\Xi_c^+\to pK^-\pi^+$ \cite{Aaij:2014yka}.
Five new $\Omega_{c}^{0}$ resonances are observed in the final
states of $\Xi_{c}^{+}K^{-}$ with $\Xi_c^+\to pK^-\pi^+$
\cite{Aaij:2017nav}. It is suggested to search for the doubly
charmed baryons in the decay of $\Xi_{cc}^{++}\to
\Xi_{c}^{+}\pi^{+}$ with $\Xi_c^+\to pK^-\pi^+$
\cite{Yu:2017zst,Wang:2017mqp}. In this work, the ratio of
fragmentation fractions \ffr~can be obtained as long as the
branching fraction of $\Xi_c^+\to pK^-\pi^+$ is determined by Eq. \eqref{eq:LHCbff}.

\section{\ffr and its implications}

Utilizing the prediction of $\mathcal{B}(\Xi_c^+\to
pK^-\pi^+)$ in Eq. \eqref{eq:Xic}, the measured value of $\mathcal{B}(\Lambda_c^+\to
pK^-\pi^+)=(6.35\pm0.33)\%$\cite{Patrignani:2016xqp} and the
reasonable theoretical ratio
$\mathcal{B}(\Xi_b^0\to\Xi_c^+\pi^-)/\mathcal{B}(\Lambda_b^0\to\Lambda_c^+
\pi^-)\approx1$, the ratio of the fragmentation fraction for $b$
quark into $\Xi_b^0$ and $\Lambda_b^0$ can be obtained from
Eq.\eqref{eq:LHCbff} as
\begin{align}\label{eq:myff}
\frac{f_{\Xi_b^0}}{f_{\Lambda_b^0}}=0.054\pm0.020.
\end{align}
The uncertainty is mainly from the branching fraction of $\Xi_c^+\to
pK^-\pi^+$ in Eq. \eqref{eq:Xic}.
This result of \ffr~is much smaller than the naive estimation of
$f_{s}/f_{u}$ or 0.2 in Eq. \eqref{eq:fsfu}, and the MIT bag model for
the branching fraction of $\Xi_b^-\to\Lambda_b^0\pi^-$ in Eq.
\eqref{eq:MIT}. The central value of our result in Eq.
\eqref{eq:myff} is one half of those obtained via
$\Xi_{b}^{-}(\Lambda_{b}^{0})\to J/\Psi \Xi^{-}(\Lambda)$ in Eqs.
\eqref{eq:voloshin} and \eqref{eq:hsiao}.
Only the prediction in the diquark model for
$\Xi_b^-\to\Lambda_b^0\pi^-$ in Eq. \eqref{eq:diquark} is
consistent with our result within the uncertainties, while the central value is larger as well.

The total $b$-baryon fraction can be obtained by the ratio \ffr~in Eq. \eqref{eq:myff}. 
The production of $\Omega_{b}^{-}$ is doubly suppressed by  two strange quarks, estimated as 15\% of the $\Xi_{b}$ \cite{Aaij:2016avz}. It is smaller than the
error of \ffr~in Eq. \eqref{eq:myff}, and thus can be neglected in the
total fraction of $b$-baryons. We then have
\begin{align}
f_{\rm baryon}=f_{\Lambda_b}+2f_{\Xi_b}+f_{\Omega_b}
\approx f_{\Lambda_b}+2f_{\Xi_b} = (1.11\pm0.04)f_{\Lambda_b}.
\end{align}
It is equivalent that the correction $\delta$ in Eq.
\eqref{eq:bbaryon} is $\delta=0.11\pm0.04$, which is smaller than
the estimation of $\delta=0.25\pm0.10$ in Ref.
\cite{Aaij:2016avz}.

With the result of \ffr~in Eq. \eqref{eq:myff}, the branching
fraction of $\Xi_b^-\to \Lambda_b^0\pi^-$ can be determined from
Eq. (\ref{eq:StrangeWD}),
\begin{align}
\mathcal{B}(\Xi_b^-\to \Lambda_b^0\pi^-)=(1.06\pm0.54)\%.
\end{align}
This is consistent with the diquark model, but larger than the MIT
bag model, seen in Eq. \eqref{BRstrangeWD}.

The precision of our result of \ffr~in Eq. \eqref{eq:myff} can be
significantly improved by the measurements of $\Xi_c^+\to p\overline
K^{\ast0}$ and $\Lambda_c^+\to\Sigma^+K^{\ast0}$ at LHCb, BESIII and Belle II. 
The large uncertainty of $\mathcal{B}(\Xi_c^+\to pK^-\pi^+)$ in Eq. \eqref{eq:Xic}, inducing the major uncertainty of \ffr, is dominated by two ratios of branching fractions: $\mathcal{B}(\Xi_c^+\to p\overline
K^{\ast0})/\mathcal{B}(\Xi_c^+\to pK^-\pi^+)=0.54\pm0.10$ by FOCUS in 2001
\cite{Link:2001rn} and
$\mathcal{B}(\Lambda_c^+\to\Sigma^+K^{\ast0})/\mathcal{B}(\Lambda_c^+\to\Sigma^+\pi^{+}\pi^{-})=0.078\pm0.022$ measured by FOCUS in 2002
\cite{Link:2002zx}.  For the former, a more precise measurement can be performed by LHCb with partial wave analysis. At LHCb with the data of 3.3 fb$^{-1}$, there are already $1\times10^{6}$ events of $\Xi_c^+\to pK^-\pi^+$
 \cite{Aaij:2017nav}, which is four orders of
magnitude larger than 200 events in Ref. \cite{Link:2001rn}. The latter can be improved by the BESIII or
Belle(II) experiments, which have recently performed a dozen 
measurements of $\Lambda_{c}^{+}$ decays
\cite{Pal:2017ypp,Ablikim:2016vqd,Ablikim:2017iqd,Zupanc:2013iki,Ablikim:2016mcr,Ablikim:2015prg,Ablikim:2015flg,Yang:2015ytm,Ablikim:2016tze,Ablikim:2017ors},
especially the observation of some singly or doubly
Cabibbo-suppressed processes
\cite{Yang:2015ytm,Ablikim:2016tze,Ablikim:2017ors}.  With the
updated measurements of $\Xi_c^+\to p\overline
K^{\ast0}$ and $\Lambda_c^+\to\Sigma^+K^{\ast0}$ in the near future,
\ffr~could be of higher precision.

%%%%%%
\section{summary}

In this work, we study the ratio of fragmentation fractions
\ffr~with the data of $\Xi_{b}^{0}\to \Xi_{c}^{+}\pi^{-}$
and $\Lambda_{b}^{0}\to \Lambda_{c}^{+}\pi^{-}$, 
$\Xi_{c}^{+}/\Lambda_{c}^{+}\to p K^{-}\pi^{+}$  at LHCb, which is the most precise measurement related to the fragmentations of $\Xi_{b}$ and $\Lambda_{b}$, seen in Table. \ref{tab:list}. The least known branching
fraction of $\Xi_{c}^{+}\to p K^{-}\pi^{+}$ is obtained under the
$U$-spin symmetry,  $\mathcal{B}(\Xi_{c}^{+}\to p
K^{-}\pi^{+})=(2.2\pm0.8)\%$. The ratio \ffr~is then determined to
be \ffr=$0.054\pm0.020$. This is the first analysis of \ffr~using
the LHCb data. It helps to understand the production of
$b$-baryons. To improve the precision, we suggest to measure the
ratios of branching fractions
$\mathcal{B}(\Lambda_c^+\to\Sigma^+K^{\ast0})/\mathcal{B}(\Lambda_c^+\to\Sigma^+\pi^{+}\pi^{-})$
and $\mathcal{B}(\Xi_c^+\to p\overline
K^{\ast0})/\mathcal{B}(\Xi_c^+\to pK^-\pi^+)$ at BESIII, Belle(II)
and LHCb using the current data set or in the near future.

\vskip 1.2cm \acknowledgments
This research was supported by the National Natural Science
Foundation of China under the Grant No. 11505083 and U1732101.

\end{document}